\documentclass[twocolumn,superscriptaddress,floatfix,preprintnumbers,amsmath,amssymb]{revtex4}

\usepackage{graphicx}% Include figure files
\usepackage{dcolumn}% Align table columns on decimal point
\usepackage{bm}% bold math
\usepackage{float}
\usepackage{natbib}
\setcitestyle{square}

\usepackage{amsmath}
\usepackage{amssymb}
\usepackage[usenames]{color}

\newcommand{\labeq}[1]{\begin{equation} #1 \end{equation}}

\newcommand{ \mb }{ \mathbf } 
\newcommand{ \qq }{ \mb{q}}

\newcommand{ \lam }{ \lambda }

\newcommand{\ket}[1]{| #1 \rangle}
\newcommand{\bra}[1]{\langle #1 |}

\begin{document}

%\preprint{APS/123-QED}

\title{How do defects limit the ultrahigh thermal conductivity of BAs? A first principles study}
%\title{Effect of substitutional defects on the thermal conductivity of Boron Asenide: a first principles study}
%\title{The thermal conductivity penalty of doping BAs}

%\title{Effect of substitutional defects on the thermal conductivity of large mass-ratio binary compounds: a Boron Arsenide case study}

\author{Mauro Fava}
 \thanks{These two authors contributed equally}
 \affiliation{LITEN, CEA-Grenoble, Grenoble, 38054, France}
 \email{Mauro.FAVA@cea.fr}
 
\author{Nakib Haider Protik}
 \thanks{These two authors contributed equally}
 \affiliation{John A. Paulson School of Engineering and Applied Sciences, Harvard University, Cambridge, MA 02138, USA}
 \email{nakib@seas.harvard.edu}

\author{Chunhua Li}
 \affiliation{Department of Physics, Boston College, Chestnut Hill, MA 02467, USA}
 \email{licf@bc.edu}

\author{Navaneetha Krishnan Ravichandran}
 \affiliation{Department of Mechanical Engineering, Indian Institute of Science, Bangalore, India}
 %\email{?@bc.edu}

\author{Jes\'{u}s Carrete}
 \affiliation{Institute of Materials Chemistry, TU Wien,Vienna, A-1060, Austria}
 %\email{?@institution.edu}

\author{Ambroise van Roekeghem}
 \affiliation{LITEN, CEA-Grenoble, Grenoble, 38054, France}
 %\email{?@institution.edu}
 
\author{Georg K. H. Madsen}
 \affiliation{Institute of Materials Chemistry, TU Wien,Vienna, A-1060, Austria}
% \email{?@institution.edu}

\author{Natalio Mingo}
 \affiliation{LITEN, CEA-Grenoble, Grenoble, 38054, France}
 \email{natalio.mingo@cea.fr}
 
\author{David Broido}
 \affiliation{Department of Physics, Boston College, Chestnut Hill, MA 02467, USA}
% \email{?@bc.edu}

\date{\today}

\begin{abstract}
The promise enabled by BAs high thermal conductivity in power electronics cannot be assessed without taking into account the reduction incurred when doping the material. Using first principles calculations, we determine the thermal conductivity reduction induced by different group IV impurities in BAs as a function of concentration and charge state. We unveil a general trend, where neutral impurities scatter phonons more strongly than the charged ones. $\text{C}_{\text{B}}$ and  $\text{Ge}_{\text{As}}$ impurities show by far the weakest phonon scattering and retain BAs $\kappa$ values of over $\sim$ 1000 $\text{W}\cdot\text{K}^{-1}\cdot\text{m}^{-1}$ even up to high densities making them ideal n-type and p-type dopants. 
Furthermore, going beyond the doping compensation threshold associated to Fermi level pinning triggers observable changes in the thermal conductivity. This informs design considerations on the doping of BAs, and it also suggests a direct way to  determine the onset of compensation doping in experimental samples.
 \end{abstract}

\maketitle

Increased power densities in electronic devices lead to heightened efficiency and durability issues due to overheating, prompting the search for alternative, higher thermal conductivity ($\kappa$) materials. Thus, the recently demonstrated high $\kappa$ of cubic boron arsenide (BAs) offers great promise, particularly for power electronics. %The zincblende-structured semiconductor boron arsenide (BAs) is a novel material of growing importance, due to its high thermal conductivity. 
Measured values of $\sim$1300 $\text{W}/(\text{K}\cdot\text{m})$ have been attained at room temperature for ultra-pure samples \cite{2018Sci...361..575K,bb2c96215ab445a1bd42f7c7d145d3e0,Tian582}, in agreement with results from first principles calculations \cite{Tian582,PhysRevLett.111.025901,2017PhRvB..96p1201F}. These high values can be explained in terms of a unique collection of properties including high bond stiffness and large mass ratio between As and B atoms, in a way that reduces the scattering phase space for the phonon modes in the material. It has also been predicted that BAs should have simultaneously high room temperature electron and hole mobilities of over 1000 $\text{cm}^{2}/(\text{V}\cdot\text{s}) $\cite{PhysRevB.98.081203}. However, active applications of BAs require doping it, which could potentially destroy its advantageously large $\kappa$. Defects such as antisites, vacancies and impurities can lower the intrinsically high thermal conductivity of BAs by orders of magnitude, and n- and p-type dopants need to be identified that preferably maintain the highest possible thermal conductivity. In past works the role played by intrinsic point defects has been highlighted (Refs. \onlinecite{doi:10.1063/1.4950970}, \onlinecite{2015ApPhL.106g4105L}, \onlinecite{PhysRevLett.121.105901}), nevertheless at the present date there is no comprehensive study about the effect of external substitution atoms. Recent investigations from first principles have addressed the thermodynamic stability of many neutral and charged dopants in BAs ~\cite{chae_point_2018,doi:10.1063/1.5058134}. Among those, the ones in column IV are of great interest due to their position between the columns of B and As in the periodic table. In particular, the high p-dopability of BAs has been recently studied in Refs. \onlinecite{doi:10.1063/1.5058134} and \onlinecite{doi:10.1063/1.5062845}. Photoluminescence and electron paramagnetic resonance experiments have been used in Ref. \cite{doi:10.1063/1.5058134} along with \textit{ab initio} calculations to point out the possibility that dopants like C and Si, behaving as acceptors, might affect the BAs-conductivity by virtue of their unintentional presence in boron precursor powders and boride based compounds. Here we show how each dopant in group IV (C, Si, and Ge), in its neutral and charged forms, affects the thermal conductivity of BAs. 
This unveils a general trend, where neutral impurities reduce the thermal conductivity more strongly than charged ones. We offer an interpretation in terms of the change in orbital occupation between the original and substituted system. We also highlight the initially counter intuitive fact that, even for substitutions involving a large mass-difference value, the mass-difference scattering can be small. Finally, we show that in BAs excessive doping beyond the Fermi level pinning point activates phonon-donor scattering events, which can either slow down the decrease of thermal conductivity or cause it to plummet, depending on the type of impurity. This should be considered in future applications of BAs. Remarkably, we find that phonon scattering by $\text{C}_{\text{B}}$ and  $\text{Ge}_{\text{As}}$ dopants is exceptionally weak. As a result, high BAs $\kappa$ values can be achieved even for high defect densities. This makes $\text{C}_{\text{B}}$ and  $\text{Ge}_{\text{As}}$ impurities ideal n-type and p-type dopants.

In insulators and semiconductors, phonons are the major carriers of heat. An applied temperature gradient sets off a drift of the phonons. In steady state condition, the phonon drift is balanced by the scattering processes that produce mode re-populations, hence leading to the phonon Boltzmann Tranport Equation (BTE)~\cite{PhysRevB.53.9064,shengbte1}. The BTE can be linearised and solved self-consistently using the single mode relaxation time approximation (SMRTA) as a starting guess. By keeping only the terms linear in $\mb{\nabla{T}}$, for a given mode $\lambda \equiv (\qq,s)$ where $\qq$ is a phonon wave-vector and $s$ is a phonon branch, one can obtain the deviation from equilibrium for its phonon distribution, i.e. $n_{\lambda} = n^{0}_{\lambda} - \partial_{T}n^{0}_{\lambda}\mb{F_{\lambda}}\mb{\nabla{T}}$, where $\mb{F}_{\lambda} = \tau_{\lambda}^{0}(\mb{v}_{\lambda} + \mb{\Delta}_{\lambda} [\mb{F}_{\lambda}])$, $\tau_{\lambda}^{0}$ is the phonon relaxation time, $\mb{v}_{\lambda}$ is the phonon group velocity, $n^{0}_{\lambda}$ is the  Bose-Einstein distribution and $\mb{\Delta}$ is a linear functional of $\mb{F}_{\lambda}$. We can interpret $\mb{F}_{\lambda}$ as a vector mean free path that measures the deviation from equilibrium induced by the thermal gradient. $\mb{\Delta} \equiv 0$ corresponds to the SMRTA. In this work we consider three-, four-, and defect-mediated two-phonon scattering processes, the latter involving either isotopes or substitutional impurities. Detailed expressions for $\mb{F}_{\lambda}$, $\mb{\Delta}_{\lambda}$ and $\tau_{\lambda}^{0}$ in the three-phonon limited case are given in Ref. \cite{shengbte1}, while the extension to the four-phonons has been highlighted in Ref. \cite{article}. Here only the three-phonon scattering is included in $\Delta_{\lam}$ while the four-phonon and defect scattering are included at the SMRTA level only with an acceptable level of accuracy, as the former are dominated by Normal processes while the latter involve mostly Umklapp processes~\cite{2017PhRvB..96p1201F}.

\begin{figure}[h]
    \includegraphics[scale=0.50]{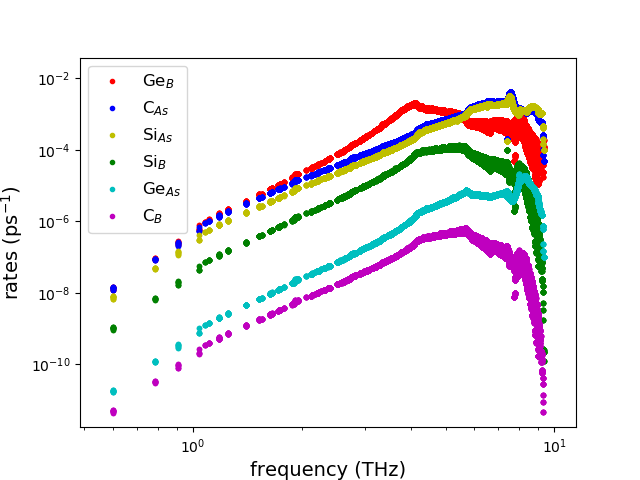}
    \caption{Phonon-defect scattering rates vs frequency, mass contribution only. The doping concentration is fixed at $\sim$ $10^{18}$ $\text{cm}^{-3}$.}
    \label{fig:mass_scat}
\end{figure}

A substitutional defect introduces an on-site mass perturbation, $V_{M}$, and an extended bond perturbation, $V_{K}$ (see Supplementary Materials). The total perturbation is $V = V_{M} + V_{K}$. Once this is known, we can calculate the retarded Green's function for the unperturbed system, $G^{+}_{0}$, and the scattering T-matrix \cite{PhysRevB.81.045408,economou1983green}. We assume that the defects are randomly distributed throughout the crystal and that their concentration is sufficiently low that each defect can be treated as an independent scattering center. Since the defect concentrations considered here are dilute, less than 0.1$\%$ of either As or B atoms, effects associated phonon scattering from multiple defects can be ignored \cite{Walton1975}. Thus, the phonon-defect scattering rates can be expressed as:

%\labeq{\label{eq:rates}
%    \dfrac{1}{\tau_{\lam}^{\text{ph-def}}} = -\chi\dfrac{\Omega_{s}}{\omega_{\lam}V_{\text{host}}}\Im\bra{\lam}T^{+}\ket{\lam}}

\labeq{\label{eq:rates}
    \dfrac{1}{\tau_{\lam}^{\text{ph-def}}} = -\chi\dfrac{V_{\text{uc}}}{\omega_{\lam}}\Im\bra{\lam}T^{+}\ket{\lam}}

where $\chi$ is the volume concentration of impurities in the system, $\omega_{\lam}$ is the angular frequency and $V_{\text{uc}}$ is the unit cell volume. Phonon transport in BAs is peculiar in the sense that it is dominated by phonons in a narrow window of frequencies between 4-8 THz. This is due to several features in BAs including a large frequency gap between acoustic and optic phonons, a narrow optic phonon bandwidth, a bunching together of acoustic phonon branches and exceptionally weak four-phonon scattering. This combination of features gives rise to unusually large contributions to $\kappa$ from acoustic phonons in this particular range \cite{PhysRevLett.111.025901,Tian582}.

Let us first look at the scattering produced by the mass difference. The absolute mass difference normalized by the host atom mass is shown in table~\ref{tab:massdef}. $\text{Ge}_\text{As}$ and $\text{C}_\text{B}$ substitutions correspond to the smallest mass differences, since
Ge and As and C and B are contiguous in the periodic table. For the rest of the substitutions the mass difference is large.

\begin{table}[h]
\centering
\begin{tabular}{c|c|c|c}
    & C & Si & Ge\\
    \hline
 B & $0.1$ & $1.6$ & $5.7$ \\
 As & $0.8$ & $0.6$ & $0.03$ \\
\end{tabular}
\caption{Absolute mass difference normalized by the host atom mass.}
\label{tab:massdef}
\end{table}
\twocolumngrid

In contrast with the case of single-species compounds, the magnitudes of the scattering rates in Fig.~\ref{fig:mass_scat} do not follow a monotonic behavior with respect to the normalized mass difference: in binary compounds with a large mass ratio of the constituent atoms, like BAs, almost all acoustic phonon modes throughout the Brillouin zone involve dominant motion of the heavy As atoms, while the light B atoms remain relatively stationary. As a result, mass defects placed on the heavy atom sites lead to strong phonon scattering while those placed on the light atom site become almost invisible to phonons and provide only weak scattering \cite{lindsay_phonon-isotope_2013}. An approximated analytical expression for the mass-difference scattering rate in large mass-ratio binary compounds was given in 
Ref.~\onlinecite{lindsay_phonon-isotope_2013}, and experimentally verified in Ref.~\onlinecite{chen_ultrahigh_2020}.

\begin{figure}
    %\centering
\includegraphics[scale=0.5]{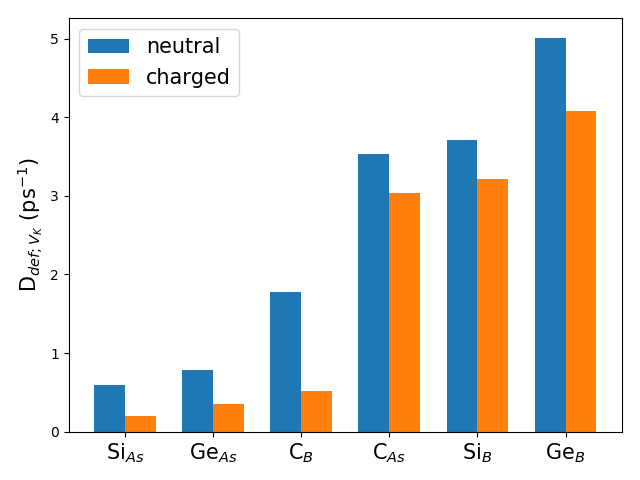}
\caption{$V_{K}$ - descriptor for charged and neutral rates.}
\label{fig:descriptor}
\end{figure}

%\begin{figure}[t!]
%    \centering
%    \begin{subfigure}[]{0.50\textwidth}
%        \centering
%        \includegraphics[scale=0.60]{RATES_BAs_NEUTRAL_ALL_2x2x2.png}
%        \caption{neutral case}
%    \end{subfigure}
%    ~
%    \begin{subfigure}[]{0.50\textwidth}
%        \centering
%        \includegraphics[scale=0.60]{FINAL_PLOTS/rates_charged.png}
%        \caption{charged case}
%    \end{subfigure}
%    \caption{$V = V_{M}+V_{K}$}
%\label{fig:rates_vk}
%\end{figure}

To assess the importance of $V_{K}$ we find useful to define and evaluate the following descriptor, for both charged and neutral states:

%\begin{equation}
%    D_{\text{def};K} \equiv \int_{\omega_{\text{min}}}^{\omega_{\text{max}}}[\tau^{-1}_{V_{tot}}(\omega) - \tau^{-1}_{V_{M}}(\omega)]d\omega
%\end{equation}

\begin{equation}
    D_{\text{def};K} \equiv \frac{1}{N}\sum_{\lambda}\tau^{-1}_{\lambda;K}[\theta(\omega_{\lambda}-\omega_{1}) - \theta(\omega_{\lambda}-\omega_{2})],
\end{equation}

%The effect of also including the bond perturbation, $V_K$, is seen in Fig.~\ref{fig:rates_vk}.
%Panels (a) and (b) show the neutral and charged impurity cases. 

where N is the number of $\qq$ - points in the reciprocal space grid, $\omega_{\text{1}}$ and $\omega_{\text{2}}$ are 4 and 8 THz, respectively, $\theta$ is the Heaviside step-function and $\tau^{-1}_{\lambda;K}$ is the phonon-defect scattering rate evaluated by including the $V_{K}$ perturbation only. The evaluated descriptor is shown in Fig.~\ref{fig:descriptor} for dopants in the neutral and charged states. The latter correspond to those with the lowest formation energy according to Refs.~\onlinecite{chae_point_2018,doi:10.1063/1.5058134}: $\text{Si}^{-1}_\text{As}$, $\text{Ge}^{-1}_\text{As}$, $\text{C}^{-1}_\text{As}$, $\text{Si}^{+1}_\text{B}$, $\text{Ge}^{+1}_\text{B}$ and $\text{C}^{+1}_\text{B}$.

\begin{figure}[ht!]
    %\centering
\includegraphics[scale=0.50]{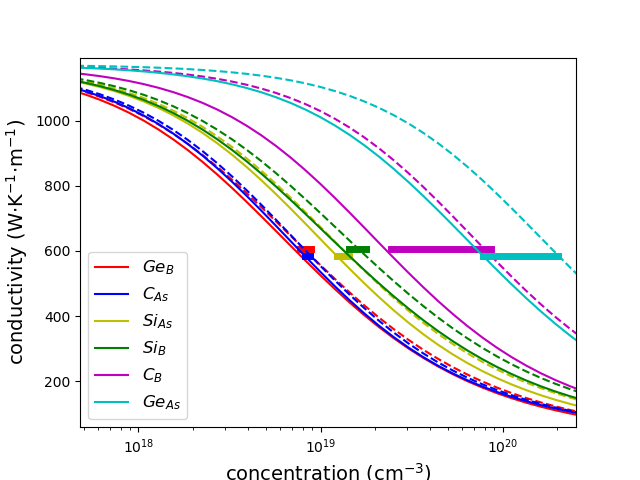}
\caption{Lattice thermal conductivity of BAs vs doping concentration at 300K. Solid lines: neutral impurities. Dotted lines: charged impurities.}
\label{fig:all_curves_300}
\end{figure}

We find that phonon scattering by neutral defects is generally stronger than that by charged defects. We can possibly attribute this general phenomenon to the fact that the ionised charged states more closely resemble the electronic structure of the original host: when an atom in column IV replaces an As (B) atom, it tends to get charged by accepting (donating) an extra electron, thus becoming iso-electronic with the original atom for which it has substituted. If the impurity remains neutral, however, the extra hole (electron) present at the defect site is responsible for bond perturbations on the crystal structure that do not take place when dealing with ionised states. This hypothesis would deserve further investigation in some future publication. This effect is most noticeable in $\text{Ge}_\text{As}$, $\text{C}_\text{B}$ and $\text{Si}_\text{As}$, as shown in Fig.~\ref{fig:descriptor}. These three impurities also have weaker bond-defect scattering than do $\text{Ge}_\text{B}$, $\text{Si}_\text{B}$ and $\text{C}_\text{As}$. This is also clear in the plots of thermal conductivity as a function of doping concentration, in Fig.~\ref{fig:all_curves_300}. Solid lines correspond to the neutral impurity cases, while dotted lines are for the charged defects. The behaviour of the conductivity clearly reflects the behaviour of the rates: the reduction of $\kappa$ is larger for the neutral impurities. The $\text{C}_\text{As}$ and $\text{Ge}_\text{B}$ defects are the impurity that affect the thermal transport the most, with a 50\% reduction to $\sim  600 \ \text{W}/(\text{K}\cdot\text{m})$ at $\chi\sim 10^{19}$ $\text{cm}^{-3}$; $\text{Si}_\text{As}$ closely follows, because its large mass variance dominates over the otherwise weak bond defect scattering. Since Si and C are known contaminants in BAs growth \cite{doi:10.1063/1.5058134}, the present finding that they strongly reduce the BAs $\kappa$ motivates synthesis approaches that minimize their presence, if maximum thermal conductivity is desired. Interestingly, $\text{Si}_\text{B}$ does provide quite a large reduction of $\kappa$ in both the neutral and charged cases despite having relatively weak low frequency rates, because the scattering induced by the bond perturbation is quite high (similar to the $\text{C}_\text{As}$ case) between 4 and 8 THz. This effect clearly cannot be captured by a pure mass perturbation. The opposite holds for $\text{Ge}_\text{As}$: it gives the smallest reduction to the BAs thermal conductivity, even in the neutral case, where phonon-impurity scattering rates are about an order of magnitude smaller than those for the $\text{C}_\text{As}$ substitution in the critical 4-8 THz range.

\begin{figure}[ht!]
    %\centering
\includegraphics[scale=0.50]{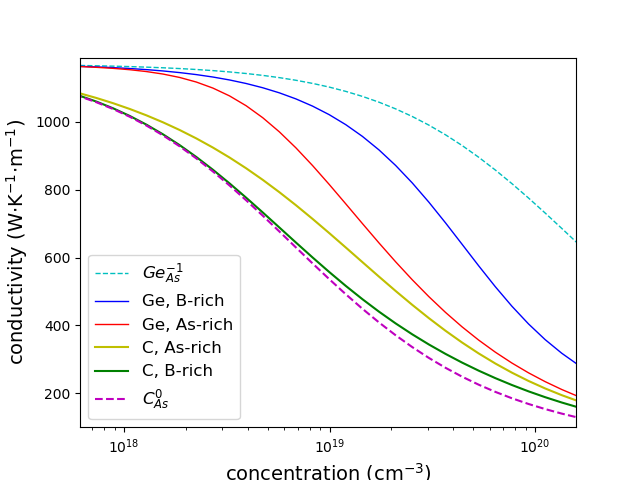}
\caption{Lattice thermal conductivity of BAs vs impurity concentration at 300K with neutral acceptors and donor compensation.}
\label{fig:compensating_300}
\end{figure}

In the C$_\text{As}$, Ge$_\text{B}$ and Si$_\text{As}$ cases the mass variance hides the effect produced by the structural relaxation and by the change in the local bonds to such degree that the difference between neutral and charged states cannot be appreciated at the $\kappa$-level, especially in the Si$_\text{As}$ case where the bond perturbation is already weak, see Fig.~\ref{fig:all_curves_300}. Instead, the effect of $V_{K}$ plays a much more relevant role when the mass variance is weak, that is in the $\text{Si}_\text{B}$, $\text{C}_\text{B}$ and  $\text{Ge}_\text{As}$ cases. In particular, in the latter two it constitutes the largest component of the perturbation and the difference between neutral and charged states is more marked. Moreover, for the $\text{Si}_\text{B}$ substitution, relatively weak rates at low frequency ($\sim < 4$THz) in the charged case are compensated by a much stronger perturbation at high frequency.

The curves of Fig.~\ref{fig:all_curves_300} ignore any thermodynamic considerations as to whether the doping levels in the abscissa are achievable for each dopant. For instance, although neutral group-IV impurities in BAs always have higher formation energies than the charged ones, a certain fraction of neutral impurities can still form at finite temperature as the Fermi level gets closer to the pinning point. Indeed, given the fraction of thermally activated acceptors $f(T)=[1 + 4\exp((\epsilon_{\text{A}}-\epsilon_{\text{F}})/k_{B}{T})]^{-1}$,  where $\epsilon_{\text{A}}$ and $\epsilon_{\text{F}}$ are the acceptor energy and Fermi Level respectively, the effective phonon-acceptor scattering rates for a system that can accommodate either kind would turn into $1/\tau_{\lambda,\text{tot}}^{\text{def}}(\chi)\sim-\chi\Im[f\bra{\lambda}{T^{+}(\text{ch.})}\ket{\lambda}+(1-f)\bra{\lambda}T^{+}(\text{ne.})\ket{\lambda}]$. Therefore understanding how a large difference between the neutral and the charged defect-induced perturbations could shape the temperature dependence of the scattering rates and thermal conductivity should be considered.
The dopability of these impurities has been studied in Refs.~\onlinecite{chae_point_2018},~\onlinecite{doi:10.1063/1.5058134}. According to Figs.~3 and 4 in Ref.~\cite{chae_point_2018}, all three dopants, D= \{C, Si, Ge\}, would start behaving as acceptors upon increasing concentration and progressively lowering the Fermi Level. At a given threshold value, $\epsilon_{\text{F}}$ becomes pinned at the value corresponding to the intersection between the D-acceptor and corresponding D-donor formation energy curves. This is the point at which acceptor and donor concentrations become equal and the effective carrier concentration cannot be increased. As the pinning energy is approached, the impurity can substitute on both As and B sites as D$^{-1}_{As}$ and D$^{+1}_{B}$, such that $\epsilon_{\text{F}}$ stays pinned even if the impurity concentration is increased further. Such donor compensation can change the thermal conductivity dependence versus impurity concentration, in light of the generally different magnitudes of the phonon-defect scattering rates for As and B substitutions. In the case of Ge doping, at lower densities Ge$^{-1}_{As}$ defects reduce $\kappa$ only slightly. But once Ge$^{+1}_{B}$ starts to form, $\kappa$ will decrease more rapidly due to the much larger scattering rates of the latter. The opposite behavior occurs for C doping since C$^{-1}_{As}$ scatters phonons more strongly than C$^{+1}_{B}$. The actual donor concentration depends on the impurity formation energies, on band structure-related quantities and on the experimental growth conditions as well - thus its precise \textit{ab initio} calculation is challenging. Nevertheless, we employed the formation energy curves from inspection of Figs.~3-4 of Ref.~\cite{chae_point_2018}, and we took the following rough pinning $\epsilon_{\text{F}}^{*}$-values (eV) for As-rich (B-rich) conditions: C: 0.4 (0.15), Si: 0.15 (-0.05), Ge: 0.25 (0.0). We used these values along with averaged effective masses $m_{h}^{*}\sim $ 0.56 and $m_{e}^{*}\sim $ 0.4 (\cite{doi:10.1063/1.5058134}) to estimate the concentrations of charged and neutral acceptors and of compensating donors for each impurity density applying the charge neutrality condition at an assumed growth temperature of 1163K, consistent with measurement. The procedure is described in the Supplementary Materials. By including the scattering rates for charged and neutral acceptors and the compensating donors at the transport temperature of 300K, (see Supplementary Materials), we have calculated the BAs $\kappa$ vs. impurity concentration curves in fig.~\ref{fig:compensating_300}. Solid lines show the Ge doping and C-doping cases for B-rich and As-rich growth conditions. The change of dependence in the case of Ge doping is particularly evident. When grown in B-rich conditions, BAs could in principle be p-doped with Ge to nearly 10$^{19}$ cm$^{-3}$, without much decrease in the thermal conductivity, with somewhat smaller but still high $\kappa$ values achieved for the As-rich case. Beyond this doping level, the thermal conductivity would decrease more rapidly compared with the Ge$^{-1}_{As}$ - only case, with detrimental consequences for device heating. We note that As-rich conditions are known to be more favorable for growth, whereas growth of B-rich BAs can be hampered by formation of the subarsenide phase $\text{B}_{6}\text{As}$ ~\cite{doi:10.1063/1.5058134}.

In terms of dopability, amongst the three dopant species, Si can achieve the highest hole concentrations. However, it also has a higher conductivity reduction, where $\kappa$ reduces to half its bulk value already at concentrations of $\sim10^{19}$ cm$^{-3}$. For the experimentally achievable As-rich growth, Ge p-doping shows a clear advantage over C and Si doping as donor compensation starts to be important only at concentrations approaching $\sim10^{19}$ cm$^{-3}$.
This can could be a great advantage in applications where efficient heat dissipation is crucial. Furthermore, the results in Fig.~\ref{fig:compensating_300} suggest a complementary experimental way to determine if compensation doping is occuring by directly measuring its thermal conductivity. Further calculations and experiments may be envisaged in this way, to evaluate the effect of compensation on the thermal transport properties in semiconductors. Finally, we note that actual growth techniques may overcome the unfavorably large formation energy of the C$_\text{B}$ impurity. The ability to create charged C$_\text{B}$ substitutions in the absence of C$_\text{As}$ would give only minimal reduction of the BAs - $\kappa$ up to high densities, and simultaneously provide a currently unavailable n-type dopant.

In conclusion, we have evaluated the thermal conductivity reduction induced by doping BAs with C, Si and Ge. Si doping can achieve higher hole densities, but its associated reduction in thermal conductivity is high. Conversely, Ge$_\text{As}$ and C$_\text{B}$ substitutions give only small reductions to the BAs - $\kappa$ even at high densities, making them particularly attractive p- and n-type dopants for device applications. An observable drop (enhancement) in thermal conductivity with respect to the charged D-acceptor case is predicted and explained upon Ge(C)-doping if we consider compensating donor scattering centers and the temperature/doping dependence for the acceptor activation. This imposes practical limitations to be considered when designing BAs - based devices. It also suggests a direct alternative way to experimentally determine if a sample suffers from compensation doping. Finally, we have computationally identified a general phenomenon whereby charged impurities isoelectronic with the substituted species scatter phonons noticeably more weakly than their corresponding neutral counterparts. This phenomenon deserves further investigation in other systems with thermodynamically stable neutral impurities.

\section*{ \label{sec:ack} Acknowledgement }

This work was supported in part by the Office of Naval Research under MURI grant No. N00014-16-1-2436, and the Agence Nationale de la Recherche through project ANR-17-CE08-0044-01. GKHM acknowledges funding from the Austrian Science Funds (FWF) under project CODIS (Grant No. FWF-I-3576-N36). We thank Nebil Katcho for providing us with the first version of the code used to compute the phonon-defect scattering rates.
%\nocite{*}

\bibliography{ref}% Produces the bibliography via BibTeX.

\end{document}